\documentclass[oldversion]{aa}
\usepackage{epsfig}
\usepackage{natbib}
\usepackage{lscape}
\usepackage{amssymb}
\usepackage{epstopdf}
\usepackage[caption=false]{subfig}

\usepackage{longtable}
\usepackage{color}
\usepackage{fancyheadings}
\bibpunct{(}{)}{;}{a}{}{,}

\begin{document}

\title{Can stellar activity make a planet seem misaligned?}


\author{M. Oshagh\inst{1,2}, S. Dreizler\inst{1}, N. C. Santos\inst{2,3}, P. Figueira \inst{2}, A. Reiners\inst{1}}

\institute{
Institut f\"ur Astrophysik, Georg-August-Universit\"at,
Friedrich-Hund-Platz 1, 37077 G\"ottingen, Germany
\and
Instituto de Astrof\' isica e Ci\^encias do Espa\c{c}o, Universidade do Porto, CAUP, Rua das Estrelas, PT4150-762 Porto, Portugal 
\and 
Departamento de F{\'i}sica e Astronomia, Faculdade de Ci{\^e}ncias, Universidade do
Porto,Rua do Campo Alegre, 4169-007 Porto, Portugal\\
email: {\tt moshagh@astro.physik.uni-goettingen.de}
}

\date{Received XXX; accepted XXX}

\abstract {Several studies have shown that the occultation of stellar active regions by the transiting planet can
generate anomalies in the high-precision transit light curves, and these anomalies may lead to an inaccurate
estimate of the planetary parameters (e.g., the planet radius). Since the physics and geometry
behind the transit light curve and the Rossiter-McLaughlin effect (spectroscopic transit) are the same, the
Rossiter-McLaughlin observations are expected to be affected by the occultation of stellar active regions in a
similar way. In this paper we  perform a fundamental test on the spin-orbit angles as derived by Rossiter-McLaughlin measurements, and we examine the impact of the occultation of stellar active regions by the transiting planet on the spin-orbit angle estimations.  Our results show that the inaccurate estimation on the spin-orbit angle  due to stellar activity  can be quite significant (up to $\sim 30$ degrees), particularly for the edge-on, aligned, and small transiting planets. Therefore, our results suggest that the aligned transiting planets are the ones that can be easily misinterpreted as misaligned owing to the stellar activity. In other words, the biases introduced by ignoring stellar activity are unlikely to
be the culprit for the highly misaligned systems.}



\authorrunning{M. Oshagh et al.}
\maketitle
\section{Introduction}


As a star rotates, the part of its surface  that rotates toward the observer will be 
blue-shifted and the part that rotates away will be red-shifted. During the transit of a planet,
the corresponding rotational velocity of the portion of the
stellar disk that is blocked by the planet is removed from
the integration of the velocity over the entire star, creating
the radial velocity (RV) signal
which is known as the Rossiter-McLaughlin (RM) effect \citep{Rossiter-24, McLaughlin-24}. This effect has been used to determine
the projected rotation velocity of a star ($v \sin i$), and
the angle between the sky-projections of the stellar spin
axis and the planetary (or the eclipsing binary's secondary) orbital plane (hereafter spin-orbit angle)\footnote{The planet-to-star radius ratio can also  be estimated through RM measurements; however, this parameter has been frequently derived from the photometric transit light curve.} \citep[e.g.,][]{Winn-05, Hebrard-08, Winn-10, Simpson-10, Triaud-10,
Hirano-11, Albrecht-12, Hirano-16}. In the era of high-precision RV measurements, like those provided by HARPS and HARPS-N and the upcoming spetrograph ESPRESSO, determination of the spin-orbit angle and $v \sin i$ can be influenced
by second-order effects such as the convective blueshift \citep{Shporer-11, Cegla-16}, the differential stellar rotation
\citep{Albrecht-12}, and the microlensing effect due to the transiting planet's mass \citep{Oshagh-13c}. 

It has been shown in several studies that the occultation of stellar active regions (i.e., stellar spots and plages) by the transiting planet can generate anomalies in the high-precision transit light curves and may lead to an incorrect
estimate of the planetary parameters \citep[e.g.,][]{Sanchis-Ojeda-11a, Oshagh-13b, Sanchis-Ojeda-13, Barros-13, Oshagh-15a, Oshagh-15b}. The detection of these anomalies in the transit light curves becomes the norm in the exoplanet community after   the high-precision photometric observations achieved by space-based telescopes (such as CoRoT and Kepler).

Since the physics and geometry behind the transit light curve and RM effect
are the same, they are both expected to be affected
by the occultation of active regions by the transiting planet in a similar way. Probing the impact of the occultation of stellar active regions anomalies in the high-precision RM observations on the spin-orbit angle measurements, and quantifying its influence is the main objective
of this paper \footnote{There have been several studies that assess the impact of non-occulted stellar active regions on the transit light curves and the planetary parameters estimations \citep[e.g.,][]{Czesla-09,Ioannidis-16}. The non-occulted stellar active regions would also affect RM measurements; however, probing this effect is beyond scope of this paper, but will be pursued in
forthcoming publications. }. In Sect. 2 we present the details of our models that are used
to produce mock RM observations. In Sect. 3 we discuss to what extent stellar activity can affect spin-orbit angle measurements. In Sect. 4, we present the possible interpretation of our results, and we conclude our study
by summarizing the results in Sect. 5.

\begin{figure}
\includegraphics[width=95mm, height=140mm]{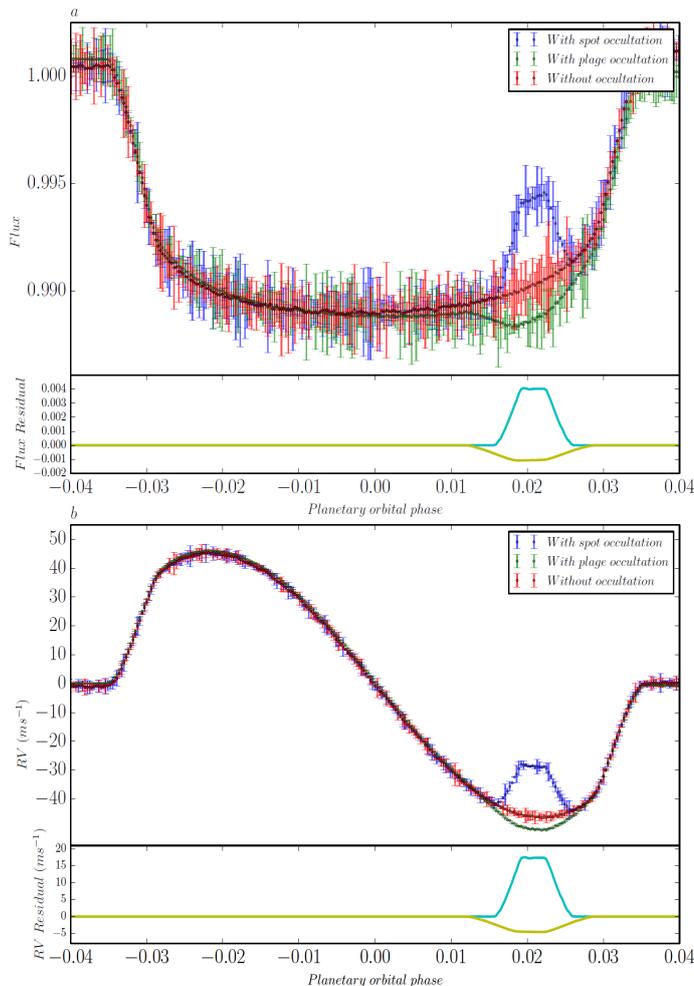}
 \caption{Top:   Simulated transit light curve anomaly due to occultation of a transiting planet with a spot or a plage shown by the blue  and green dots, respectively. The red dots show the simulated transit light curve without occultation with any active regions. The simulation was done for an aligned Jupiter-size transiting planet overlapping a dark spot with filling factor of 0.5\% and plage with filling factor of 2.25\%. The error bar obtained by adding random Gaussian
noise at the level of 300 ppm (which is consistent with the photometric precision achievable by Kepler for an 11 magnitude star on short cadence).
Bottom: Same as the top panel for the simulated RM observations. The error bar obtained by adding random Gaussian
noise at the level of  1 $m s^{-1}$ (which is consistent with the RV precision achievable by current
spectrographs such as HARPS and HARPS-N). }
  \label{sample-figure}
\end{figure}


\section{Generating  mock RM observations}

In this study we use the publicly available tool SOAP2.0 \citep{Dumusque-14}. This tool has the capability of generating the photometric and radial velocity variation signals induced by active regions on a rotating star by taking into account not only the flux contrast effect in those regions, but also the RV shift due to inhibition of the convective blueshift inside those regions. To be able to simulate the RM observations we modified the SOAP2.0 tool in order to include a transiting planet in the system while taking into account the overlap between the transiting planet and active regions; the same procedure was performed on the SOAP tool \citep{Boisse-12} to develop the SOAP-T tool \citep{Oshagh-13a}. The modified SOAP2.0 (hereafter  SOAP2.0-T) allows us to generate the transit light curves, RM signals, and the occultation anomalies inside the transits and RMs. We note that the main difference between SOAP-T and SOAP2.0-T is that SOAP2.0-T takes into account the inhibition of the convective blueshift inside active regions, which affects the RV measurements, as was shown in \citet{Dumusque-14}. Figure 1 presents examples of transit light curves and RM observations, which both show occultation anomalies generated by the SOAP2.0-T.

\subsection{Model parameters}
To start generating our mock RM observations sample, first we have to pinpoint the proper parameters of the systems to be simulated. We considered the central star to be a solar-like star, a G-dwarf with the temperature of $T_{eff} = 5778 K$.  We assigned the stellar quadratic limb-darkening coefficients  
$u_{1}= 0.29$ and $u_{2} = 0.34$, which correspond to quadratic limb-darkening coefficients of a star with a temperature close to that of the Sun ($\sim  5800 K$)
\citep{Sing-10, Claret-11}. We assume that the stellar rotation axis is parallel to the plane of sky (edge-on stellar rotation). We assigned two different stellar rotation rates for the star,  $v \sin i=3$ and $10 kms^{-1}$.

We chose the filling factor\footnote{The filling factor can be defined as $f = (R_{spot}/R_{\star})^{2}$, where $R_{spot}$ is the spot radius.} of the stellar spots in our simulations to be
 1\%, which corresponds to the largest filling factor
of sunspots \citep{Solanki-03}. For
the stellar plages, we considered f = 6.25\%, the maximum filling factor
of the Sun's plages \citep{Meunier-10}. Both spots and plages are considered to have circular shape in the SOAP2.0-T tool. Several studies have demonstrated that by assuming circular  active regions they manage to model perfectly the observed transit light curve anomalies due to occultation of active regions  \citep[e.g.,][]{Sanchis-Ojeda-11b, Sanchis-Ojeda-13, Oshagh-13a}; therefore, we consider the circular shape assumption to be  realistic.


We used the reported value for the maximum temperature contrast of the stellar spot on the solar-like star which is $\Delta T_{spot}= - 2000 K$ \citep{Berdyugina-05}. We used the temperature contrast of the plage as suggested by \citet{Meunier-10} and as it is implemented in SOAP2.0-T,
\begin{equation}
\Delta T_{plage}=250.9-407.7\cos{\theta}+190.9\cos^2\theta,
\end{equation}
where $\theta$ is the angle between the normal to the stellar surface and the observer.  We note that the plages show a limb brightening behavior that is  opposite to the s spots', which show limb darkening \citep{Meunier-10}. We  note that the maximum filling factor and temperature contrast of the Sun's active regions normally occur during the maximum of the Sun's magnetic activity cycle \citep{Solanki-03}.

\begin{figure}[t!]
\includegraphics[width=95mm, height=65mm]{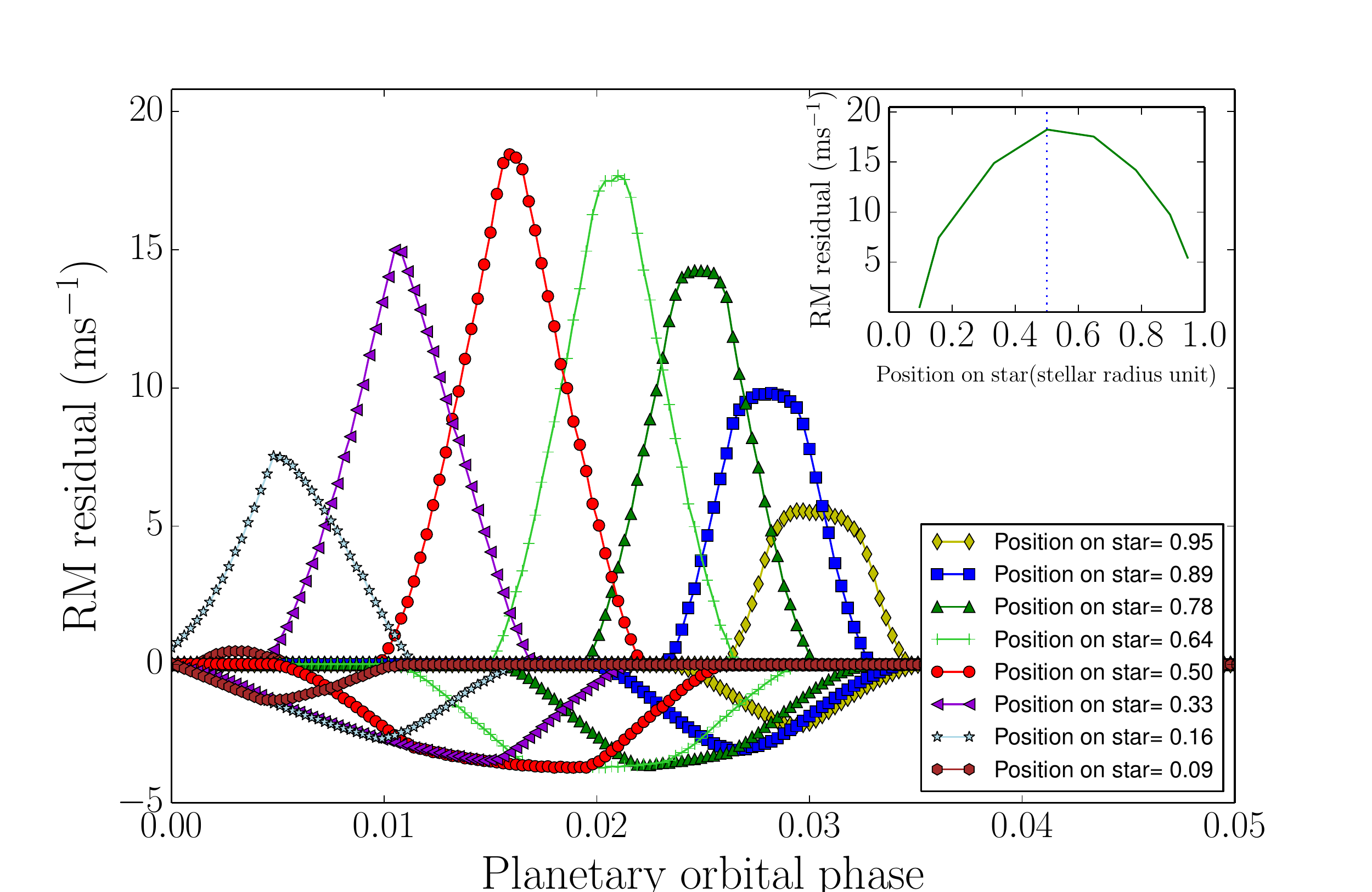}
 \caption{The amplitude of RM anomalies as a function of the time that anomaly appears in the RM signal. RM anomalies for the case of occultation with stellar spot are shown in the positive part of y-axis and the occultation with the stellar plage are presented in the negative part of y-axis. Panel on the top-right corner
 shows the same but as a function of the position of planet-active region occultation (0 on the x-axis means overlap occurs
while the active region is at the center of stellar disk, and 1 means occurs on the limb of the stellar disk).}
  \label{sample-figure}
\end{figure}

We chose a transiting planet on a three-day period orbit.
We took the planets to be of Jupiter (J) and
Neptune (N) sizes. The planet-to-star radius ratios in these systems are 0.035 and 0.1 for a Neptune-sized planet transiting a G-dwarf and for a Jupiter-sized planet transiting a G-dwarf,
respectively. The main reason behind choosing these system is their large values
of planet-to-star radius ratio, which makes them favorable for the detection of their RM signal. We assigned the impact parameter of transiting planet ($b$) to be 0 and 0.5. We also considered  three different spin-orbit angles for the transiting planet as $\lambda= 0, 30,$ and $60$ degrees.

\subsubsection{When exactly does the maximum RM anomaly occur?}
Since we are interested in the configurations that increase the effect and maximize the amplitude of the RM anomaly, we need to identify the longitudes of the active regions that maximize this effect. To this end, we need to determine the exact moment when the overlap between the transiting planet and
the stellar activity region occurs, generating the maximum of the RM anomaly. To do this we performed the following test.

We considered an aligned transiting planet with a radius of $\frac{R_{p}}{R_{\star}}=0.1$ with zero impact parameter. We assumed a dark spot with a filling factor of 0.5\% located on the stellar equator. We assigned all these parameters to the SOAP2.0-T tool, and generated the synthetic RM observations with anomalies. In order to  obtain the RM residual we also generated synthetic
RMs with the same initial conditions but without
considering the occultation of the active region by planet. The subtraction
between the RM without occultation anomaly and the RM with anomaly due to active region occultation yields the RM anomaly signal (see Figure 2). We repeated this process for different
values of the spot's longitude and determined the amplitude
of the RM anomaly as a function
of the anomaly position in the RM observations. We repeated the same test for a case of a plage with a filling factor of 6.25\% also located on the stellar equator. Figure 2 presents the RM anomalies for the case of occultation with a stellar spot (positive part of y-axis) and the occultation with a stellar plage (negative part of y-axis).

As the results show, in
the the top right  panel of  Figure 2, the amplitude of the RM anomalies reaches  its maximum value when the occultation occurs at the half of stellar radius (0 is the center of the stellar disk, and 1 is the limb of the stellar disk). Interestingly, the maximum RM anomaly occurs at the peak of RM signal, a region that contains the most information on spin-orbit angle. Therefore, the maximum RM anomaly   will surely have the greatest effect on the spin-orbit angle estimation. As a consequence, for all of our simulations we set the initial longitude of active regions to the values at which the occultation between the transiting planet and the stellar
activity regions will occur at 0.5 stellar radius. We list the initial longitude and latitude of active regions in the Table 1, which shows that the latitudes of active regions fall in the range of -30 and 41 degrees,   in agreement with the range of observed for the sunspot latitudes (known as the sunspot butterfly diagram).

\begin{table}[h]
\scriptsize
\caption{Initial longitudes and latitudes of active regions in our simulations}
\begin{center}
\begin{tabular}{ c c c }
\hline
\\
 Configuration & Longitude (degrees) & Latitude (degrees)\\
\\
$b=0$, $ \lambda=0$, $v \sin i=3$ & 23 & 0 \\
$b=0.5$ $ \lambda=0$, $v \sin i=3$ & 23 &-30 \\
$b=0$ $ \lambda=30$, $v \sin i=3$ & 21& 20 \\
$b=0.5$ $ \lambda=30$, $v \sin i=3$ & 24&-12 \\
$b=0$ $ \lambda=60$, $v \sin i=3$ & 10 &41 \\
$b=0.5$ $ \lambda=60$, $v \sin i=3$ & 22&-8 \\
 \\
 $b=0$, $ \lambda=0$, $v \sin i=10$ & -36 & 0 \\
$b=0.5$ $ \lambda=0$, $v \sin i=10$ & -34& -30 \\
$b=0$ $ \lambda=30$, $v \sin i=10$ & -38&20 \\
$b=0.5$ $ \lambda=30$, $v \sin i=10$ & -35&-12 \\
$b=0$ $ \lambda=60$, $v \sin i=10$ & -42&41 \\
$b=0.5$ $ \lambda=60$, $v \sin i=10$ &-35& -8 \\

\hline
\end{tabular}
\end{center}
\label{default}
\end{table}%

In summary, we considered configurations that increase the effect and maximize the amplitude of RM anomaly by considering a maximum $\Delta T$ and maximum filling factors of active regions similar to  those of the Sun during its maximum activity. However, we would like to note that recent studies have suggested that many known exoplanets orbit stars with higher levels of stellar activity than the Sun (e.g., \citealt{Basri-13} have shown that 25\%- 33\% of stars in the Kepler field of view are more active than the Sun). We would like to emphasize again that we are only considering the effects of active regions and not the other potential effects, e.g., center-to-limb convective blueshift, granulation,
oscillations, differential rotation.

\section{Measuring the spin-orbit angle}

We generated a total of 48 mock RM observations (the combination of all configurations explained in the previous section), in which all the RMs harbor occultation of the active regions' anomalies. We generated the mock RM observations with the time-sampling of 200 seconds (see Sect.4.1 where we  discuss the possible impact of time-sampling on our results). In this section we aim to probe how much the occultation of an active region's anomaly in RM
can affect the spin-orbit angle estimations. In order to do this, we fit all the generated mock RMs to estimate the spin-orbit angle and also the $v \sin i$. 

\begin{figure*}
  \centering
  \subfloat{\includegraphics[width=0.45\textwidth, height=130mm]{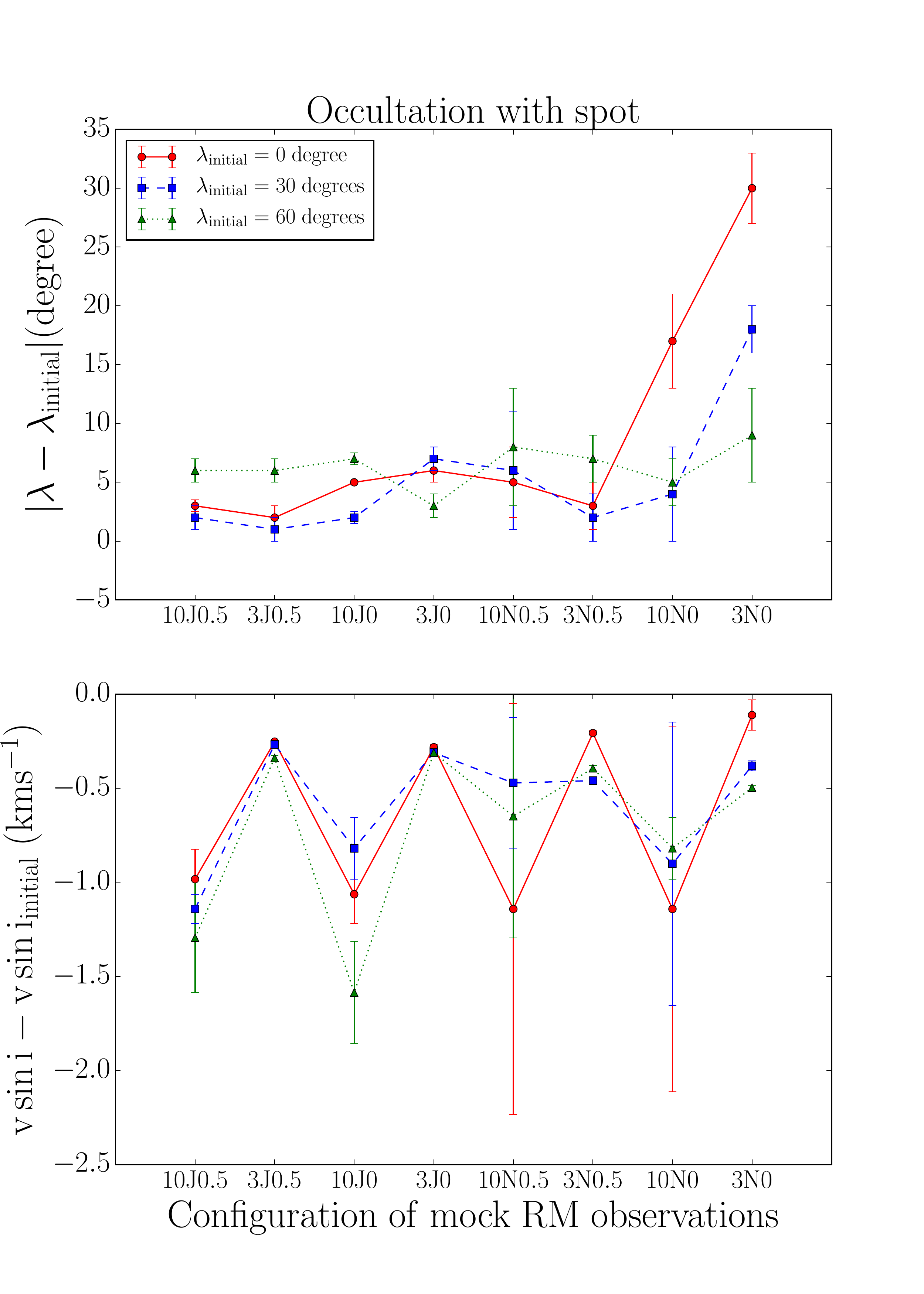}}\qquad
  \subfloat{\includegraphics[width=0.45\textwidth, height=130mm]{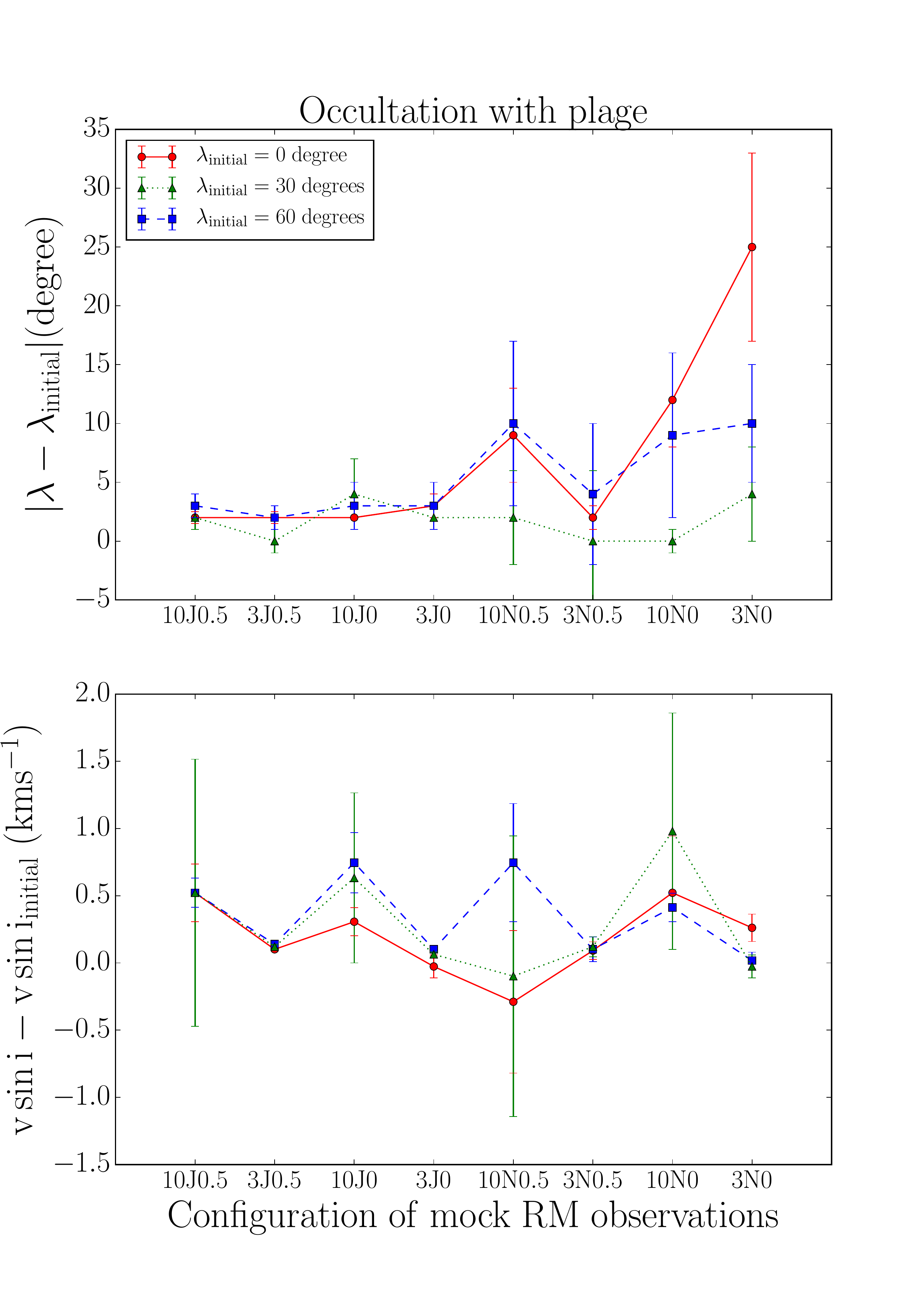}}
\caption{Top left: Deviation of the estimated value of spin-orbit angle  (obtained from the best-fit model) from its exact
value (used to generate mock RM)
as a function of different configurations. See the text for details on the x-axis description. The red solid line with the circle markers shows the cases of aligned transiting planets ($\lambda=0$ degree), the blue dashed line with square markers shows the cases of transiting planets with $\lambda=30$ degrees, and the green dotted line with triangle markers shows the cases of transiting planets with $\lambda=60$ degrees. Bottom left: Same as the top panel, but for  $v \sin i$. Top and bottom right: Same as the top left and bottom left panels, but for the RM anomalies due to occultation with the stellar plages.}
\label{fig:1}
\end{figure*}

We fit the mock RM observation with the synthetic RM of a system in which the
effect of the spot-planet overlap is not taken into account. In general, the
synthetic RM show the exact same behavior as those with anomalies,
except they do not have any RM anomalies. To make our mock RM observations closer to the real RM observation we add random Gaussian
noise at the level of 1 $m s^{-1}$ (which is consistent with RV precision achievable by current
spectrographs such as HARPS and HARPS-N) \footnote{We discuss the possible influence of RV precision on our results in Sect. 4.2.}. During the fitting procedure,
we allow the spin-orbit angle and $v \sin i$ to vary as  free parameters,
while fixing other parameters to those values used
to generate the mock RMs \footnote{In most of studies that used high-precision RM observations to estimate the spin-orbit angle of transiting planets, most of the parameters of system are known from the previous transit light curves and RV observations; therefore, the only free parameters that are estimated from the RM observations are the spin-orbit angle and $v \sin i$.}. We present two examples of mock RM observations and their best-fit  models  in  Appendix A.  The best-fit model provide us the best-fit value for the spin-orbit angel and $v \sin i$. We estimated the error bars on these fitted values
using the bootstrap method \citep{Wall-03}. 



We note that there are several analytical formulae that have been proposed to model RM signals, e.g., the ARoME tool \citep{Boue-12b}. \citet{Boue-12b} compared the ARoME tool predication with the SOAP-T's results and found a disagreement between the two. The main explanation for this difference was assuming some simplistic approximations  that were used in the analytical formulae. Because we intend to explore only the impact of the stellar active region's occultation anomaly in this study, and we do not want to be influenced
by any second-order effects,  we did not use analytical formulae to fit our mock RM observations.

Figures 3 shows the deviation of spin-orbit angle and $v \sin i$, obtained from the best-fit model and from their exact values, which were used to generate the mock RM observations. The x-axis in Figure 3 represents the configuration in which the mock RM observations were generated. The first digit stands for the $v \sin i$ in units of  $kms^{-1}$, the letter stands for the planet type, and the last digit shows the impact parameter of the transiting planet. For instance,  ``10J0.5'' means a Jupiter-sized transiting planet transits a G-dwarf star, which has $v \sin i= 10 kms^{-1}$, with the impact parameter of 0.5. Different symbols and colors show the different initial spin-orbit angles  used in generating mock RMs. The left and right panels of Figure 3  represent the results due to occultation with the stellar spot and the stellar plage, respectively. 

As shown in Figure 3, inaccurate estimation of the spin-orbit angle can be quite significant (up to $\sim 30$ degrees), particularly for the edge-on, aligned, and small transiting planets. However, the results regarding the inaccurate estimation of the $v \sin i$ do not show any particular pattern, but indicate that the under- or overestimation of $v \sin i$  can reach up to  $ 2kms^{-1}$.

These results provide some evidence that highly misaligned or retrograde Jovian transiting planets (e.g., the ones with the initial spin-orbit angle of 60 degrees; see Figure 3), are unlikely to be
affected by biases from stellar activity. In other words, the aligned transiting planets are the ones that can be misinterpreted as misaligned owing to the stellar activity. Our results can provide a viable explanation for the few cases in the literature that obtained conflicting spin-orbit angles, for instance the case of active star WASP-19 \citep{Hellier-11, Albrecht-12}.

As we noted earlier,  we assumed that
the stellar rotation axis is parallel to the plane of sky in all our simulations, and the occultation of an
active region occurs when this feature is on the half of the stellar
disk. As a result of these assumptions, the values presented here
are upper limits of the impact of RM anomaly on the spin-orbit and  $v \sin i$ estimations.

\section{Discussion}
\subsection{ Influence of time-sampling on our results}

In this section, we aim to assess the impact of time-sampling of RM observations on our results. To this end we choose one configuration in which the maximum inaccurate estimation on the spin-orbit angle was measured in Sect. 3. This system is the one with an aligned Neptune-sized planet transiting a G-dwarf with $v \sin i=3 kms^{-1}$, which in Figure 3 (left panel) is represented as ``3N0'' in the red dot. The original mock RM observation of this system was generated by time-sampling of 200 seconds. By binning the mock RM observation of this system we generate mock RM observations with the time-sampling of 600, 1000, and 1600 seconds. Then we performed the same fitting procedure, as explained and used in Sect. 3, on each RM observation. We note that  we only present the fitted value of the spin-orbit angle here (not the $v \sin i$). As shown in Figure 4, the inaccuracy on the spin-orbit angle estimation increases, up to $\sim 40$ degrees, for the cases with long time-sampling (or long exposure time).

So far, most of the RM observations have been obtained using 3-meter class telescopes equipped with a spectrograph able to deliver high-precision RV measurements. Since obtaining reasonable signal-to-noise
ratios (S/N) of a not very bright star with a 3-meters class telescope requires more than 1000 seconds exposure time, and since the transit durations are not usually longer than 3-4 hours, the number of data points observed during the RM observations is usually low (usually around 7-10 data points during each transit have been observed). This can be the explanation for why  we have not been able until now to clearly identify the RM anomalies due to occultation of active region in real RM observations, and why there has not been any attempt to correct their impact on spin-orbit angle estimations. However, in the near future, with spectographs mounted on the large telescopes, like ESPRESSO on the VLT, we will be able to reach higher S/N in a much shorter exposure time, and as a result we will be able to obtain many more data points during one transit. Therefore, the breakthrough that happened in clearly detecting the transit light curve anomalies after having space telescopes (such as Kepler and CoRoT) will happen in the RM observations after the ESPRESSO-type spectrograph.

\begin{figure}[h!]
\includegraphics[width=90mm, height=60mm]{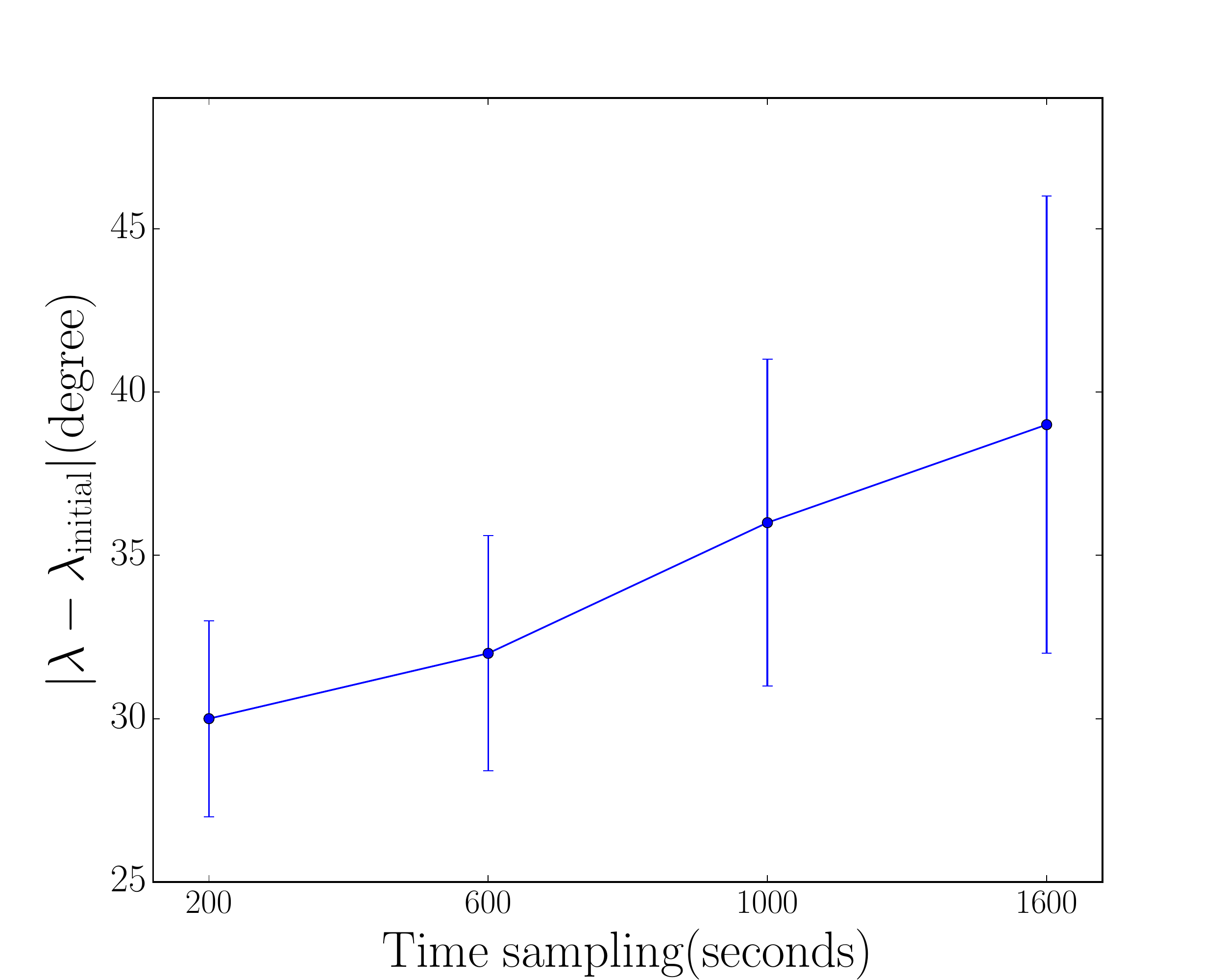}
 \caption{Deviation of the estimated value of spin-orbit angle  (obtained from  the best-fit model) from its exact
value (for the case of ``3N0'') as a function of time-sampling.
}
  \label{sample-figure}
\end{figure}

\subsection{ Impact of RV precision on our results}

In this section, we  test the influence of the RV precision on our results. Similar to Sect 4.1, we only consider the RM observation of  system ``3N0''. In Sect. 3 we added random Gaussian
noise at the level of 1 $m s^{-1}$ to the original mock RM observations of this system. To probe the impact of RV precision, here we add random Gaussian
noise at two different levels of 0.5 and 0.1 $m s^{-1}$ to the mock RM of this system. We note that the 0.1 $m s^{-1}$ RV precision is the expected RV precision that will be achieved by ESPRESSO on VLT. We again performed the same fitting procedure and measure the spin-orbit angles. As Figure 5 suggests, by improving the RV precision we will improve the accuracy on the spin-orbit angle estimation.

\begin{figure}[h!]
\includegraphics[width=90mm, height=60mm]{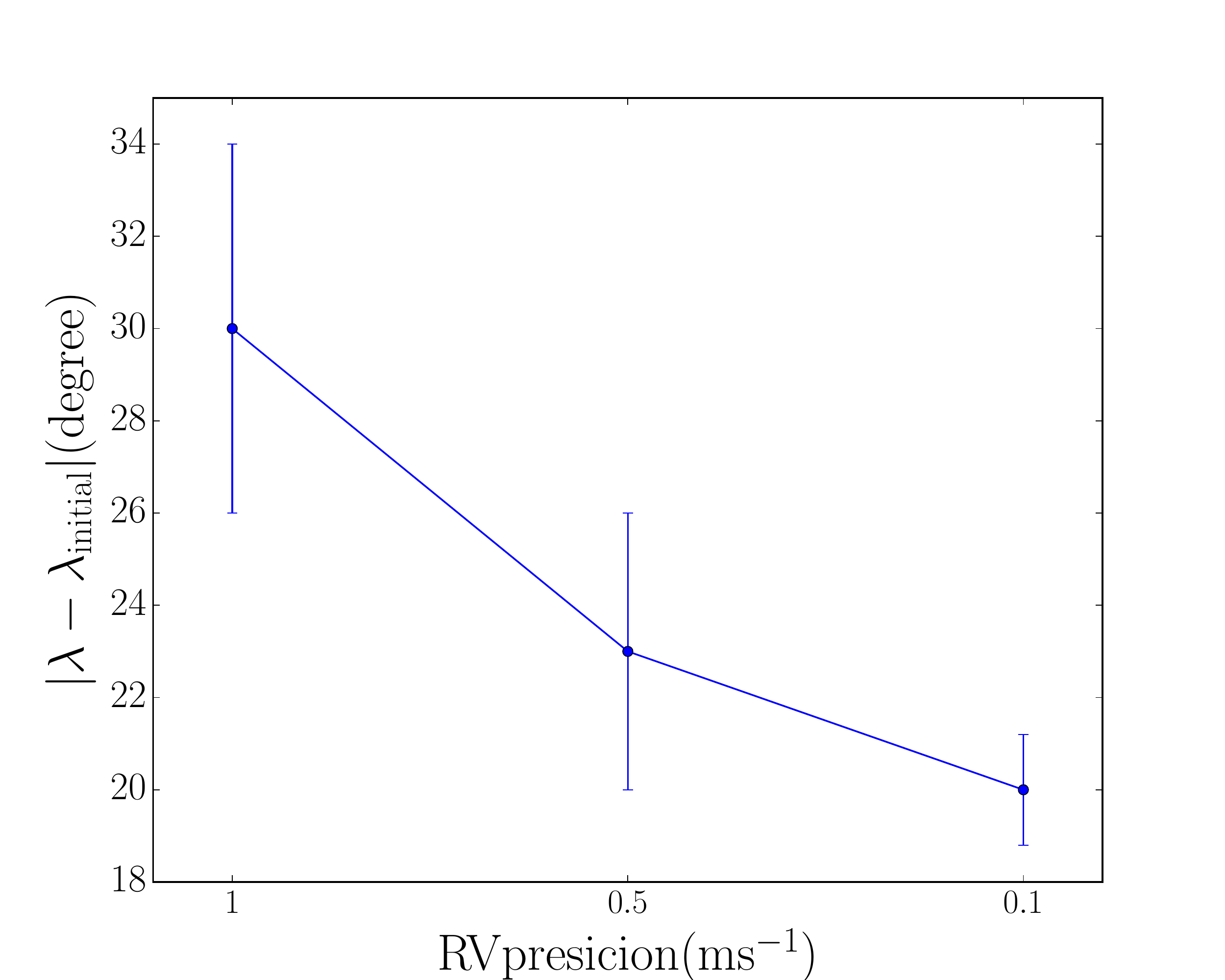}
 \caption{Deviation of the estimated value of spin-orbit angle  (obtained from the best-fit model) from its exact
value (for the case of ``3N0'') as a function of RV precision.}
  \label{sample-figure}
\end{figure}

\subsection{Probing the impact of RM observation at different wavelengths on our results}

\citet{Snellen-04} proposed the interesting idea of using RM observations in different wavelengths to obtain planet radius as a function of wavelength, and to thus retrieve the transmission spectra of the transiting planet's atmosphere. The numerical simulations performed by \citet{Dreizler-09} proved the feasibility of this technique. Recently, \citet{DiGloria-15} and \citet{Borsa-16} presented the first application of this technique on the transiting planet HD189733 b, and obtained the slope in the transmission spectrum of HD189733 b, which was interpreted as the Rayleigh
scattering in the planet atmosphere. We would like to note that the RM anomalies may also affect the planet radius estimation through the RM observations. As a consequence the RM anomaly can affect the transmission spectra measurements in the same way that the transit light curve anomaly does \citep{Oshagh-14}.

In this section we  examine the possible influences of observing RM observations at different wavelengths on our results. As we did in   Sects. 4.1 and 4.2, we only consider the RM observation of  system ``3N0''. The original mock RM observation of this system, in Sect. 2, was generated at the visible band (600 nm).

There are three wavelength-dependent
stellar parameters in the SOAP2.0-T, namely the coefficients of
quadratic stellar limb darkening ($u_{1}$ and $u_{2}$) and the relative
brightness of stellar active regions. By adjusting these parameters we will be able to produce
mock RM observations at different wavelengths. Table 2 lists
the values of $u_{1}$ and $u_{2}$ adopted from the catalog by \cite{Claret-11} for a G-dwarf as a function of wavelengths. We use the values  used in \cite{Oshagh-14} for the relative
brightness of stellar spot on a G-dwarf, which are also listed  in  Table 2. The wavelengths in  Table 2 were chosen to cover the
entire visible spectrum and the near- and mid-infrared ranges.

\begin{table}[h]

\caption{Quadratic limb-darkening coefficients of a G-dwarf and relative
brightness of stellar spots on a G-dwarf at different wavelengths \citep{Claret-11, Oshagh-14}.}
\begin{center}
\begin{tabular}{c c c c c c }
\hline
\\
$\lambda(nm)$& 400 &800 & 1500& 3000&4500\\
\\
\hline
$u_{1}$ &0.70&0.30&0.10&0.07&0.05\\
$u_{2}$ &0.18&0.30&0.32&0.14&0.12\\
Relative
brightness & 0.04 & 0.28 & 0.40& 0.58 & 0.61 \\

\hline
\end{tabular}
\end{center}
\label{default}
\end{table}%

We generate the mock RM observations of ``3N0'' at different wavelengths by considering proper values of the  limb-darkening coefficients and relative brightness of the spots at each specific wavelength. We again performed the same fitting procedure on each mock RM observation, and measured the spin-orbit angles. The results, presented in Figure 6, indicate that at long wavelengths the accuracy on the spin-orbit angle estimation will significantly improve. It is important to note that the relative
brightness of a plage on a G-dwarf does not vary as a function of wavelength as much as a spot's relative
brightness does  (as shown in \cite{Oshagh-14}; therefore, we would  expect the impact of occultation with a plage on the RM observation  not to be wavelength dependent.

This result demonstrates that measurement of RM signal in the near-infrared (NIR) regime will be less affected by the stellar activity. Therefore, we would like to suggest that the RM observations should be part of the main objective of upcoming NIR spectrographs such as CARMENES \citep{Quirrenbach-14}, the Habitable Zone
Planet Finder (HPF) \citep{Mahadevan-12}, and SPIRou \citep{Artigau-14}.

\begin{figure}[h]
\includegraphics[width=95mm, height=60mm]{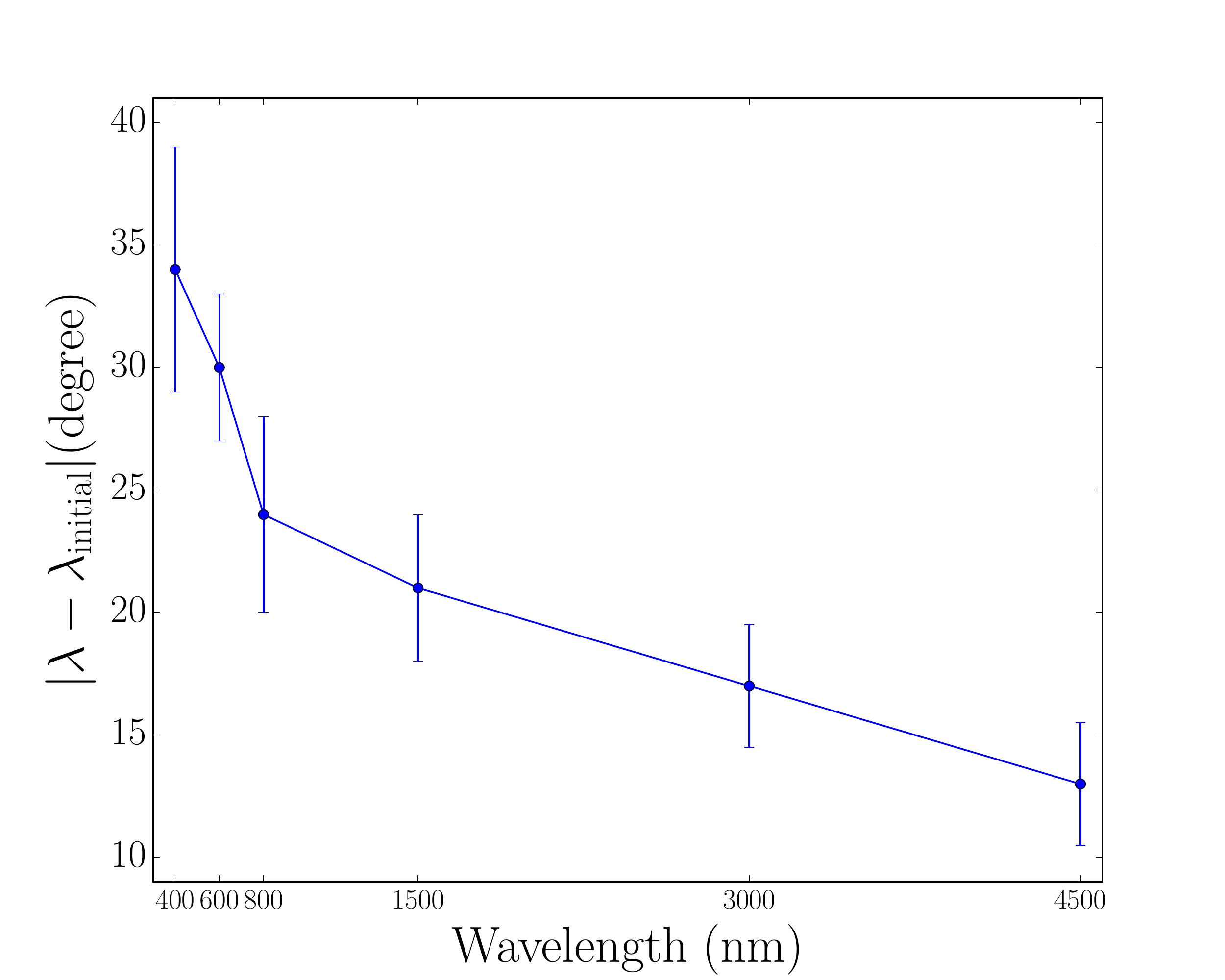}
 \caption{Deviation of the estimated value of spin-orbit angle  (obtained from the best-fit model) from its exact
value (for the case of ``3N0'') as a function of wavelength that the RM observation is performed. }
  \label{sample-figure}
\end{figure}

\section{Conclusions}
In this paper we performed a fundamental test on the spin-orbit angles as derived by RM measurements, and we examined the impact of stellar active regions occultation by the transiting planet on the spin-orbit angle estimations. We generated mock RM observations with RM anomalies for different configurations, and then analyzed them to quantify the impact of stellar activity. Our results showed the inaccurate estimation on the spin-orbit angle can be quite significant (up to $\sim 30$ degrees), particularly for the edge-on, aligned, and small transiting planets. Our results also suggested that the aligned transiting planets are the ones that can be easily misinterpreted as misaligned owing to the stellar activity. Our results can be used to explain the conflicting spin-orbit angles for the small number of  cases in the literature.

Our results also indicate that the inaccuracy on the spin-orbit angle estimation increases, up to $\sim 40$ degrees, for the cases with long time-sampling. Moreover, we discovered that increasing the RV precision will improve the accuracy on the spin-orbit angle estimation. Finally, our results suggest that performing RM observation at long wavelengths (e.g., NIR) will help to improve the accuracy on the spin-orbit angle estimation.

\begin{acknowledgements}

\scriptsize MO acknowledges research funding from the Deutsche
Forschungsgemeinschft (DFG, German Research Foundation) - OS 508/1-1. PF and NCS acknowledge support by Funda\c{c}\~ao para a Ci\^encia e a Tecnologia (FCT) through Investigador FCT contracts of reference IF/01037/2013 and IF/00169/2012, respectively, and FCT project ref. PTDC/FIS-AST/1526/2014 through
national funds and by FEDER through COMPETE2020 (ref. POCI-01-0145-FEDER-016886). PF further acknowledges support from Funda\c{c}\~ao para a Ci\^encia e a Tecnologia (FCT) in the form of an exploratory project of reference IF/01037/2013CP1191/CT0001. Last but not the least, we would
like to thank the anonymous referee for constructive comments and insightful
suggestions.


\end{acknowledgements}

\bibliographystyle{aa}
\bibliography{mahlibspot}

\newpage
\Online

\begin{appendix}



\section{Example of mock RM observations and their best-fit models }
In this section we  present two examples of the best-fit models to two of our mock RM observations, for the system  ``3N0'' with a spin-orbit angle of 0 degrees; the first example occults   a stellar spot and the second  a  stellar plage, presented in Figure A.1 and A.2, respectively. We chose these two systems because they are the ones that showed the most inaccurate estimation of the spin-orbit angle in Sect. 3.

\begin{figure}[h]
\includegraphics[width=90mm, height=70mm]{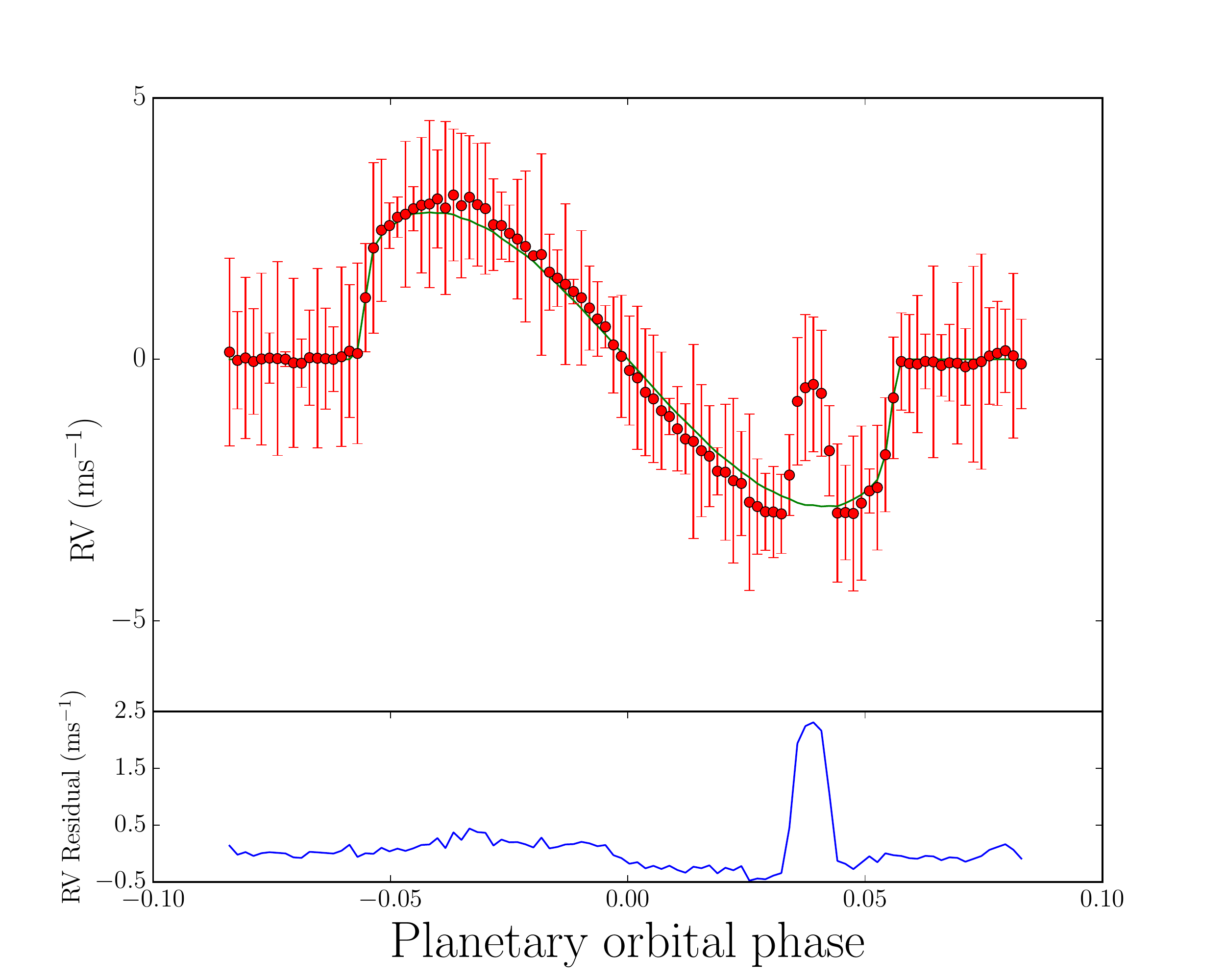}
 \caption{Mock RM observations for a system ``3N0'' with spin-orbit angle of 0 degrees. The RM anomaly is due to a stellar spot occultation. The green curve
shows the best-fit RM to the mock RMs.}
  \label{sample-figure}
\end{figure}

\begin{figure}[h]
\includegraphics[width=90mm, height=70mm]{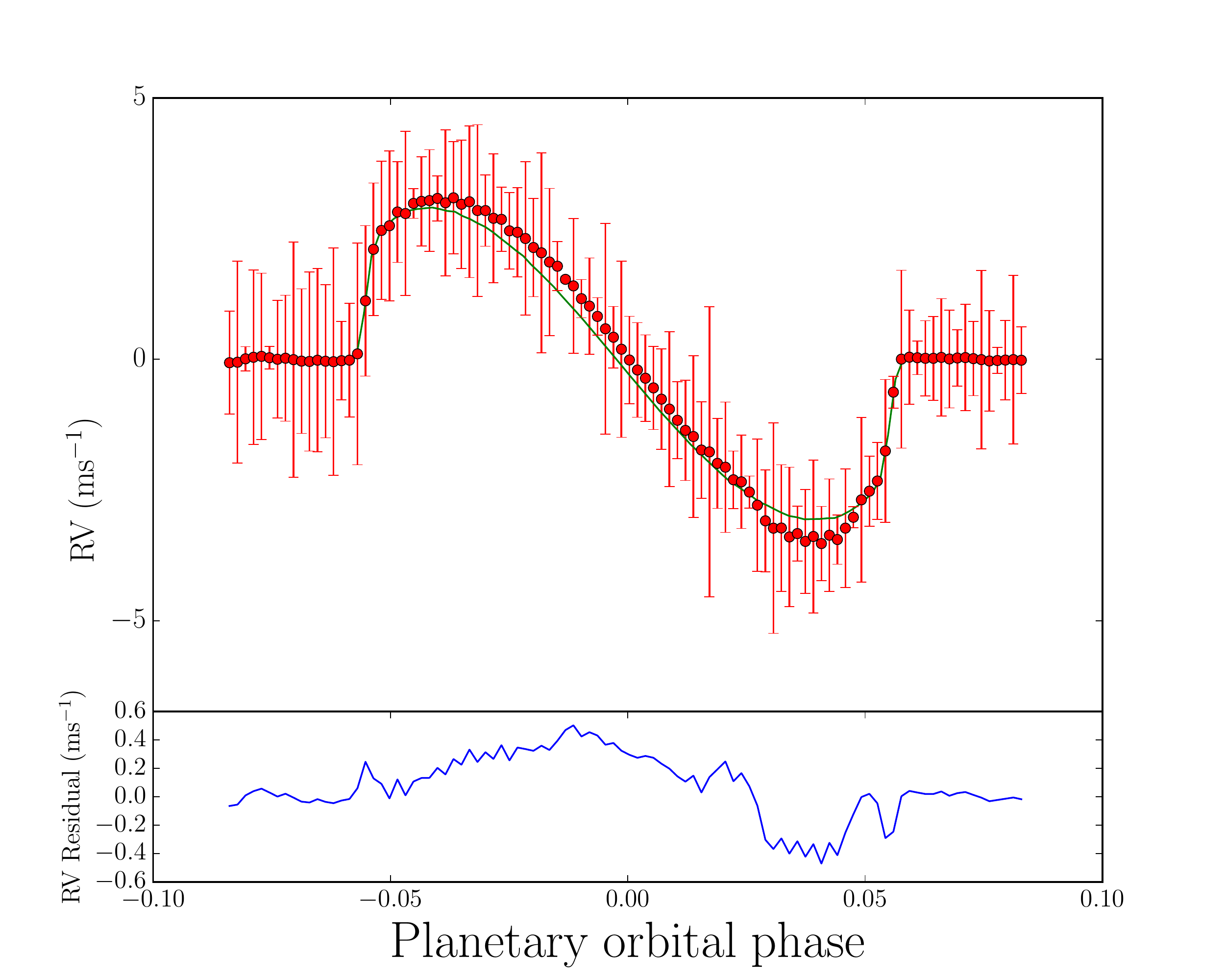}
 \caption{Mock RM observations for a system ``3N0'' with spin-orbit angle of 0 degrees. The RM anomaly is due to a stellar plage occultation. The green curve
shows the best-fit RM to the mock RM.}
  \label{sample-figure}
\end{figure}

\end{appendix}

\end{document}